\documentclass[dvips]{acta}
\usepackage{supertabular,lscape,epsfig}
\usepackage{amssymb}
\usepackage{amsmath}
\usepackage{multirow}
\usepackage{booktabs}

\newcommand{\vsini}{$v \sin i$ }

\SetPages{0}{0}

\SetVol{00}{2017}

\begin{document}

\begin{Titlepage}
\Title{Looking for the weak members of the C$_{60}^+$ family in the interstellar medium}

\Author{G~a~l~a~z~u~t~d~i~n~o~v,~G.A.}{Instituto de Astronomia, Universidad Catolica del Norte, Av. Angamos 0610, Antofagasta, Chile\\
Pulkovo Observatory, Pulkovskoe Shosse 65, Saint-Petersburg, 196140, Russia\\
Special Astrophysical Observatory, Nizhnij Arkhyz, 369167, Russia\\
e-mail: runizag@gmail.com}

\Author{K~r~e~{\l}~o~w~s~k~i,~J.,}{Center for Astronomy, Nicholas Copernicus University, Grudzi{\c{a}}dzka 5, Pl-87-100 Toru{\'n}, Poland\\
e-mail: jacek@umk.pl}

\Received{Month Day, Year}
\end{Titlepage}

\Abstract{We demonstrate, using the high resolution spectra from
the ESPADONS spectrograph, fed with the 3.6m CFH telescope, that
the strength ratios of the strong--to--weak spectral features,
attributed to C$_{60}^+$, are variable. We found that in the range of expected
9366~\AA\ C$_{60}^+$ feature there are two diffuse bands centered at
9362.0$\pm$0.1 and 9365.3$\pm$0.1 \AA\ with variable intensity ratio.
We confidently confirm the lack of 9428~\AA\ feature which, in the laboratory spectra
of C$_{60}^+$, is stronger than 9366~\AA. The weakest laboratory feature, near 9348.4~\AA\,
remains below the level of detection in all spectra. The intensity ratio 9577/9365  is variable. These facts contradict
to their common origin and so -- the identification of some interstellar
spectral features as being carried by the cation of the ``soccer ball''. We also refined the rest
wavelength position of the strongest diffuse band in this range: it is 9576.8$\pm$0.1~\AA.
}
{ISM}

\section{Introduction}

The recent publication of Campbell et al., (2015) re--started the
discussion on whether the C$_{60}^+$ (``soccer ball'') molecule may
be the carrier of two pretty strong, likely interstellar bands,
observed in near--infrared at 9577 and 9632~\AA. For a discussion
of most recent publications see Galazutdinov et al. (2017). The
latter paper demonstrated that even the ratio of the two strongest
features, attributed to $C_{60}^+$, is variable if all
possible contaminations: stellar and telluric are properly taken
care of. We could not find the weak members of the $C_{60}^+$
laboratory spectrum family but the region is severely contaminated
with telluric lines which may create doubts as to whether the
identifications or lack of them is reliable.

The publication of Walker et al.(2015) states that the weak
$C_{60}^+$ bands near 9366 and 9428~\AA\ can be traced in the spectra
of HD183143 (only 9366) and HD169454 (both). In Walker (2016) the authors reported
the detection of diffuse bands 9632, 9577, 9428, 9365 and 9348
towards the stars HD 46711, HD169454, HD183143.

The aim of this paper is to check whether the above mentioned conclusions are correct,
using the spectra from the same instrument and, as far as possible,
of the same objects. For HD183143 and HD169454 we use the same spectra, that were
anaylised by Walker et al. (2015, 2016).

\section{Spectral data}

We have downloaded four spectra from the ESPADONS archive; spectra of HD169454, BD +40\,4220 and of
CygOB2\,12 were recorded in 2005 by Bernard Foing. We have also downloaded a spectrum of HD183143, recorded in 2015 by Gordon Walker.
All four studied objects are substantially reddened hot supergiants (see Table 1).

Owing to the more than 4000 m altitude of the Mauna Kea Observatory, spectra of ESPADONS possess much weaker telluric lines
in comparison to the UVES data we have analyzed in Galazutdinov et al. (2017), thus
the range of expected weaker members of C$_{60}^+$ family can be analyzed much more confidently.
To remove the telluric lines we have downloaded spectra of HD120315 and HR7235 observed
using the same instrument. The latter object was used as a divisor in papers by Walker et al.
We found that this object shows some weak diffuse interstellar bands like 5780 and 6284 which appear in some slightly reddened spectra
(Galazutdinov et al. 1998). Theoretically, similar components of $C_{60}^+$ bands may alter their measurements.
Fortunately, HD120315 is free of any interstellar features, thus we preferred it as a divisor.

\begin{table*}
\caption{Basic parameters of the observed stars and measurements data.
There are given: interstellar rest wavelength estimates ($\lambda_0$) in~\AA, the full width at the half of maximum (FWHM) in km/s,
equivalent widths (EW) in m\AA,  equivalent widths (in m\AA) of the major diffuse bands at 5780 and 5797~\AA.
The ratio of equivalent width of diffuse bands 9577~\AA\ and 9366~\AA\ (right component) (9577/9365$_{right}$) is given in the last column of second section.
}
\label{starsir}
\scalebox{0.6}{
\begin{tabular}{cccccc ccccccccccc}
\hline

\multirow{2}{*}{Star}
             &
             \multirow{2}{*}{Sp/L}
                      &  \multirow{2}{*}{V}
                              &  \multirow{2}{*}{B-V}
                                      & \multirow{2}{*}{E(B-V)}
                                               & \multirow{2}{*}{\vsini}
                                                           &  \multicolumn{3}{c}{9577.0}   & &  \multicolumn{3}{c}{9365.2 left}    & &\multicolumn{3}{c}{9365.2 right}   \\
                                               \cmidrule{7-9} \cmidrule{11-13} \cmidrule{15-17}
             &        &       &       &        &           & $\lambda_0$ & FWHM  &  EW   & &$\lambda_0$  & FWHM   &  EW        & &  $\lambda_0$ &  FWHM & EW    \\
\hline
BD +40\,4220 &  O7Ia  &  9.13 &  1.68 & 1.97   & $>$200    & 9576.8 & 103 & 352$\pm$62   & &9361.9       & 108    & 113$\pm$20 & &9365.2 &  94 & 101$\pm$31   \\
CygOB2\,12   &  B3Iae & 11.48 &  3.22 & 3.35   &  40       & 9576.8 &  98 & 415$\pm$61   & &9362.2       & 113    & 100$\pm$20 & &9365.2 &  90 &  90$\pm$30   \\
HD169454     &  B1Ia  &  6.70 &  0.91 & 1.10   &  39       & 9577.0 & 111 & 136$\pm$46   & &9361.9       & 90     & 43$\pm$15  & &9365.6 & 112 &  43$\pm$18     \\
HD183143     &  B7Ia  &  6.84 &  1.20 & 1.24   &  37       & 9576.7 &  96 & 330$\pm$26   & &9362.1       & 85     & 30$\pm$6   & &9365.0 &  77 &  55$\pm$8    \\
\hline
HR7437       &  B7V   &  5.00 & -0.09 & 0.04   &           &        &     &              & &             &        &            & &       &     &              \\
HD120315     &  B3V   &  1.87 & -0.19 & 0.00   &           &        &     &              & &             &        &            & &       &     &              \\
\hline
\hline
   Star      &   EW(5780)   & EW(5797)    &  9577/9365right &  &           &        &     &              & &             &        &            & &       &     &              \\
\hline
BD +40\,4220 &  753$\pm$42  & 224$\pm$21  &   3.49$\pm$1.23 &  &           &        &     &              & &             &        &            & &       &     &              \\
CygOB2\,12   &  930$\pm$124 & 329$\pm$52  &   4.61$\pm$1.68 &  &           &        &     &              & &             &        &            & &       &     &              \\
HD169454     &  468$\pm$17  & 156$\pm$8   &   3.16$\pm$1.70 &  &           &        &     &              & &             &        &            & &       &     &              \\
HD183143     &  766$\pm$18  & 195$\pm$8   &   6.00$\pm$0.99 &  &           &        &     &              & &             &        &            & &       &     &              \\
\hline
HR7437       &   35$\pm$1   &   5$\pm$1   &                 &  &           &        &     &              & &             &        &            & &       &     &              \\
\hline
\end{tabular}}
\end{table*}

All measurements have been performed with the aid of DECH\footnote[1]{http://gazinur.com/DECH-software.html} code.
The wavelength scale of all spectra was aligned to the interstellar
rest wavelength position using the K{\sc i} 7699 and CH 4300~\AA~lines.
The neutral potassium profile is quite narrow in all spectra, except HD183143 (Fig. 1). In the latter it is composed of two Doppler
components, separated by about 14 km/s (Herbig \& Soderblom 1982). It is thus the problem
to which of the components the spectrum should be moved? In this paper we decided to use
for this purpose the weaker K{\sc i} component as it is evidently related to the stronger CH (4300.3~\AA)
Doppler line. This is why our central wavelength of 9577 differs in this paper from that of
Galazutdinov et al. (2017). The intensity of this band is higher in this publication because it
follows the measurement in ESPADONS spectrum which allows better removal of telluric lines (the
latter are less saturated). Anyway, both measurements coincide within the calculated errors.

Equivalent widths errors were estimated using the equation 7 from Vollmann \& Eversberg (2006) which includes contributions from
both photon noise and continuum uncertainties.

All the downloaded spectra are of very good quality, with exception in the
violet range where the polarimetric equipment makes the resultant
signal--to--noise ratio reasonably low. This is not important for
us since we try to separate infrared interstellar spectral features
from the stellar spectra, though we cannot discuss possible relations between
the considered DIBs and interstellar molecules, such as CH, CH$^+$ and CN.

\begin{figure}
\centering
\includegraphics[angle=270,width=11cm]{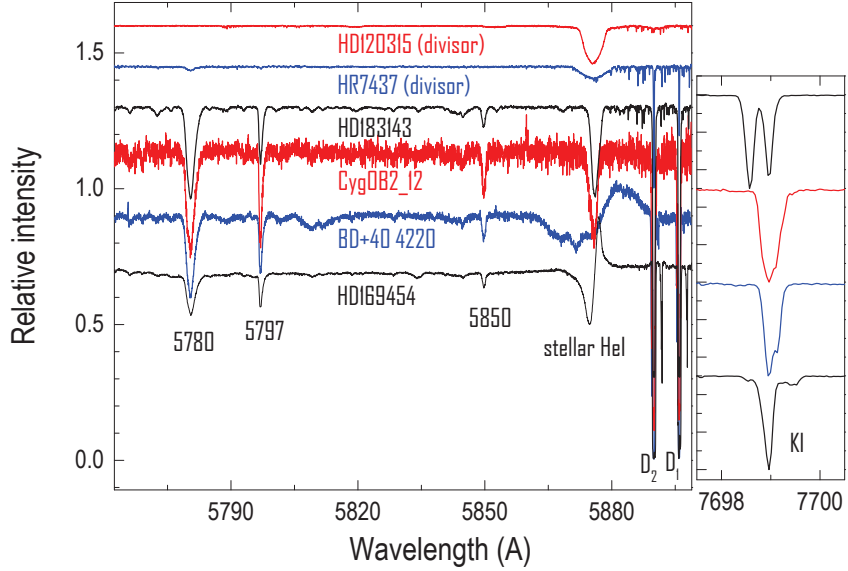}
\caption{Profiles of interstellar KI 7699~\AA\ (right) and visual spectra of reddened objects with
spectra of telluric line divisors. Note the identity of interstellar
diffuse bands while the stellar He{\sc i} is severely broadened by the fast rotation of BD +40 4220.
   }
\label{fig1}
\end{figure}

Our Fig.  1  presents the visual range of spectra of all our targets from Table 1. It is
evident that the spectrum of HD120315 is completely free of any interstellar
lines or bands. On the other hand the spectrum of HR7437, used as divisor in Walker et al.
(2015, 2016), contains the interstellar
sodium doublet as well as the weak 5780~\AA\ DIB. The other broad DIB, 6284~\AA\, is seen
as well. Apparently the target may contain some broad interstellar spectral features
(like those, attributed to $C_{60}^+$), which may introduce some uncertainties.
The variable ratio of 5780/5797 DIBs demonstrates
that the environments along the chosen sightlines are not of identical physical
parameters. It is also clearly seen that the stellar He{\sc i} line profile is
very broad and extended in the spectrum of BD +40\,4220 in contrast to other objects
(see the rotational velocities, listed in Table 1).
This is important as interstellar features' profiles remain the same in all
heavily reddened spectra but stellar ones must be evidently different which
allows to distinguish between unidentified DIBs and possible weak stellar lines.

\section{Results}

We have divided the spectra of our reddened targets by that of the selected standard.
The result is shown in Fig. 2. The remnants of telluric lines are owing to the saturation
of the atmospheric features. They can be easily removed manually. It is evident that
the 9577~\AA\ band, attributed to $C_{60}^+$, is strong in all four spectra of reddened targets and
free of even telluric remnants which makes its profile reliable. The second feature --
9428~\AA\ is doubtful. As already described by Walker et al. (2015) the expected position of
the $C_{60}^+$ feature coincides in HD183143 with a stellar emission. We identified it as the
line of Fe{\sc ii}. There is no emission
in the spectrum of the heavily reddened CygOB2\,12 but the  expected interstellar feature remains below the level of detection (see Fig. 3).

\begin{figure}
\centering
\includegraphics[angle=270,width=11cm]{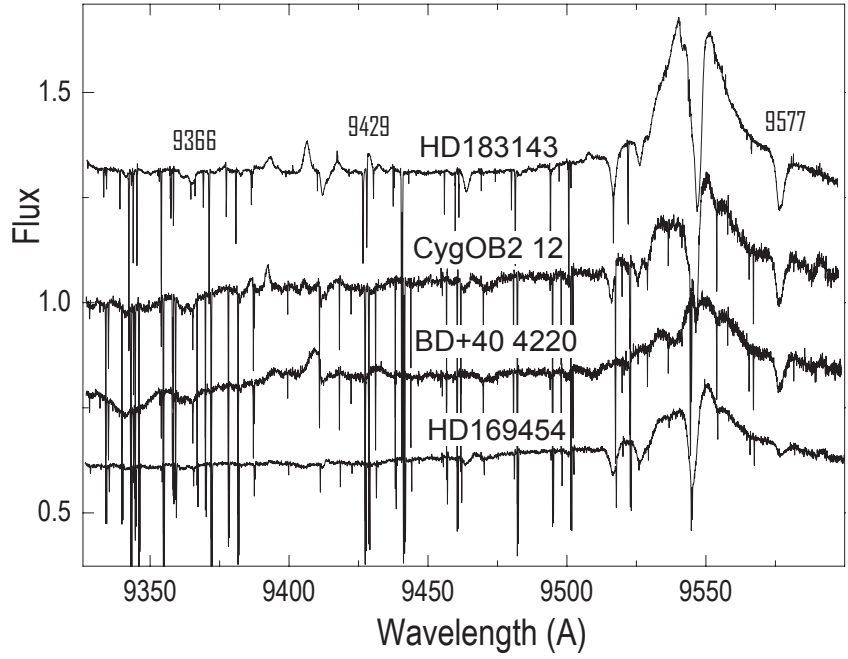}
\caption{The spectral range (divided by the telluric line divisor HD120315) of infrared features attributed to $C_{60}^+$. Note the clear
presence of the strong 9577~\AA\ DIB and the weaker 9366~\AA\. The expected 9428~\AA\ line of C$_{60}^+$ is not detectable
though in the lab spectrum it is 50\% stronger than the 9366 one.}
\label{fig2}
\end{figure}

The third possible feature -- 9366~\AA\ is present in all four spectra but its profile is
evidently ``W'' shaped. This may be either due to different physical conditions of the
intervening medium (if this is a single band) or due to the fact that we observe two DIBs, partially blended
and of another origin. We measured it as two separate diffuse bands (Table 1).
 In any case this feature is much stronger than the 9428~\AA\ one (if the
latter is present at all) which is opposite to the laboratory expectations.

\begin{figure*}
\centering
  \begin{tabular}{@{}c@{}}
    \includegraphics[width=4.38 cm, angle=270, clip=]{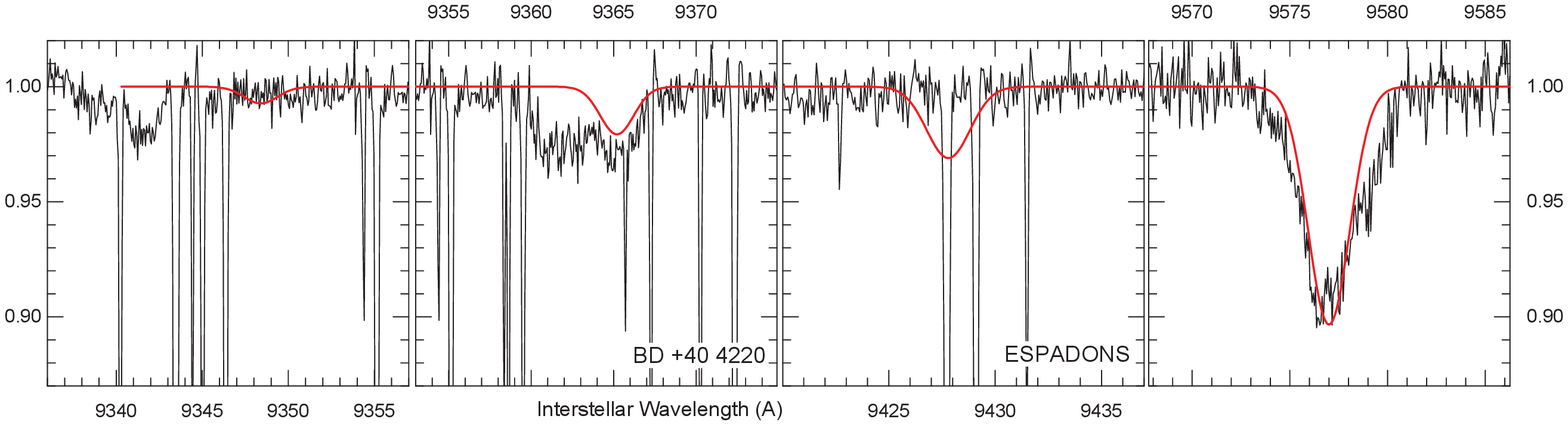} \\
    \includegraphics[width=16.5 cm,  clip=]{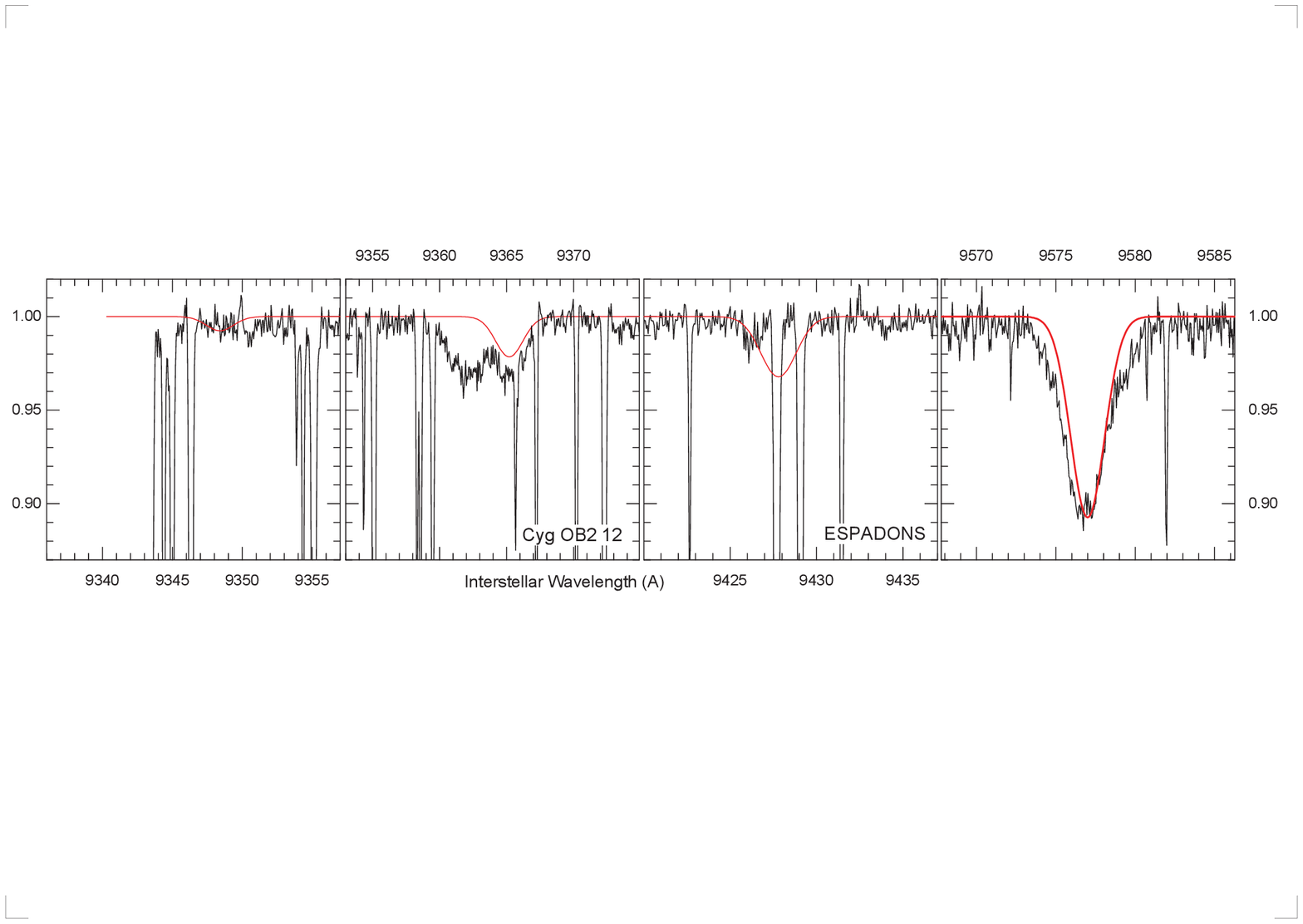} \\
    \includegraphics[width=4.4 cm, angle=270, clip=]{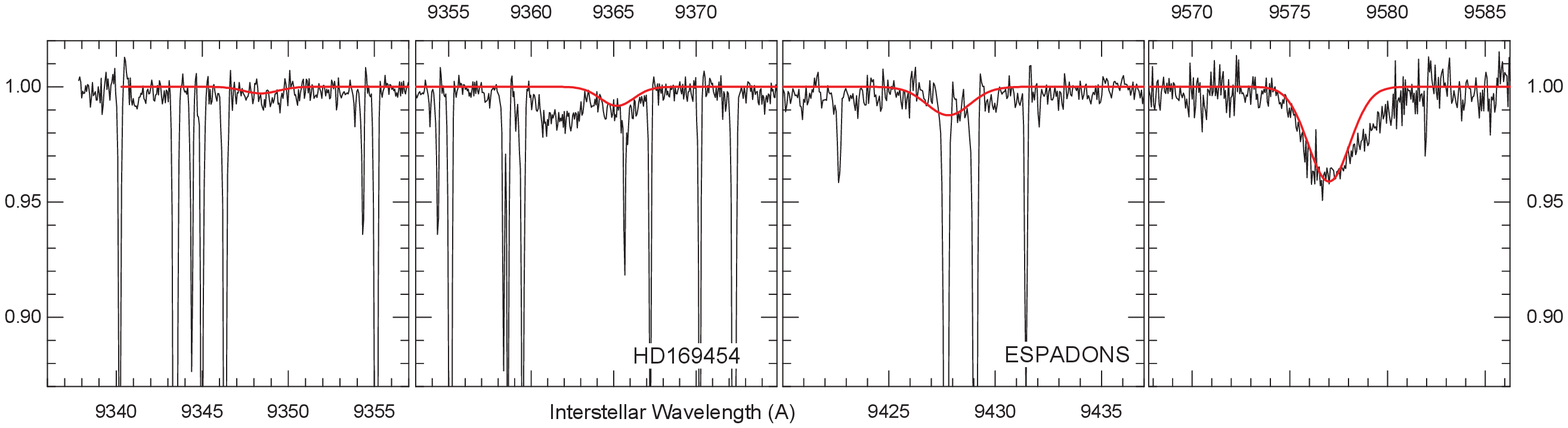} \\
    \includegraphics[width=4.5 cm, angle=270, clip=]{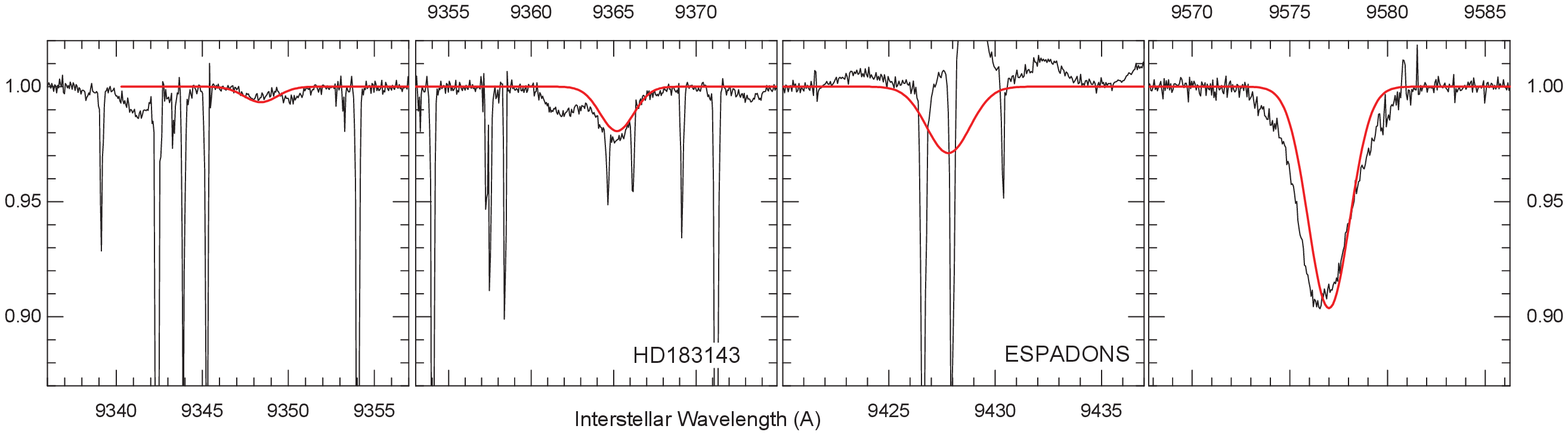} \\
  \end{tabular}
  \caption{Diffuse bands in the programme stars shown together with gaussians which mimic positions, widths and relative intensities of
the expected C$_{60}^+$ features taken from Campbell et al. (2016).  Note the absence of interstellar 9428~\AA\ feature which is the
second strong C$_{60}^+$ feature as seen in the laboratory spectrum (smooth gaussians) in the presented wavelength range.
Intensity range in all figures is the same! }
\label{allstars}
\end{figure*}

\begin{figure*}
\centering
\includegraphics[angle=270,width=15cm]{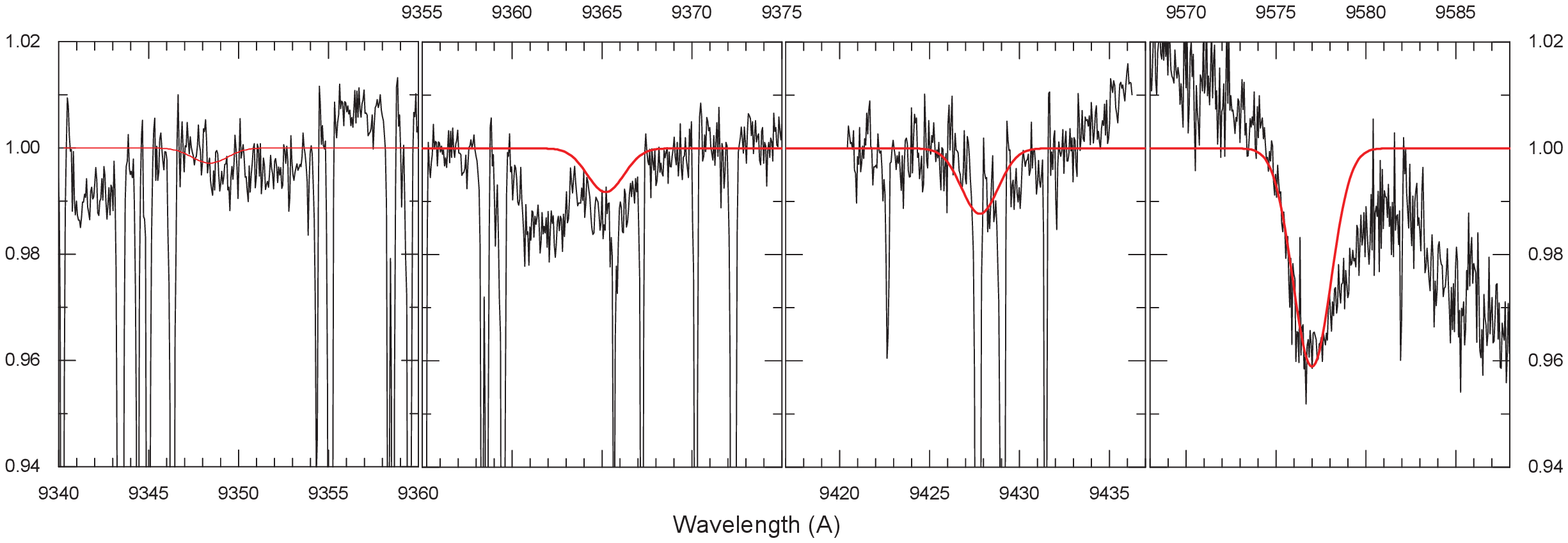}
\caption{
Spectrum of HD\,169454 after telluric lines removal with a divisor HD\,120315. No continuum normalization was made!
Smooth gaussians mimic positions, widths and relative intensities of
expected C$_{60}^+$ features, normalised to the intensity of the strong DIB9577. Note the lack of expected DIB9428~\AA\ which
should be about 50\% stronger than the observable band at 9366~\AA\ if both
these features are carried by C$_{60}^+$  (Campbell et al., 2016).
}
\label{no-contin}
\end{figure*}

Let us emphasize that our spectrum of HD\,169454 is just the same that was analyzed in Walker et al. (2016) where the authors reported the presence
of DIB 9428 being about 50\% stronger than the clearly seen 9366~\AA\ one.
However, in our Fig. 3, the lack of the 9428~\AA\ feature is surprisingly evident. Note that the continuum normalization offers no difficulties for this object.
Nevertheless, to avoid speculations about uncertainties introduced by this procedure, the spectrum of HD\,169454 is also shown right after telluric lines removal, i.e.
{\bf before} the continuum normalization (Fig. 4). For clarity, each depicted fragment of the spectrum was normalized to a constant, in order
of facilitating the comparison of all four studied
wavelength ranges.
Only the area around 9577~\AA\ exhibits some tilt (due to broad and strong hydrogen line in the spectrum of divisor) while other areas
are almost identical to those in Fig. 3 where the continuum normalization was done. Indeed, the origin of feature shown in the upper panel of
Fig. 5 of Walker et al (2016). is unclear for us.
It stands to mention that the use of HR7235 as a divisor (the same divisor was used in Walker et al., 2016) exhibits similar result,
i.e. the DIB 9428~\AA\ cannot be revealed (Fig. 5).

\begin{figure}
\centering
\includegraphics[angle=270,width=8cm]{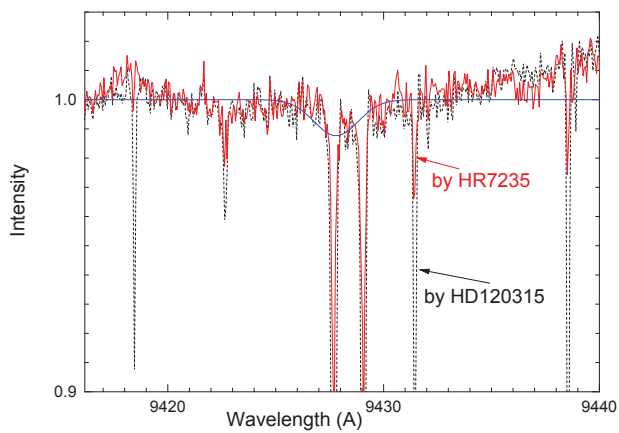}
\caption{
Spectrum of HD\,169454 after telluric lines removal by HR7235 (solid red line) and, by HD\,120315 (dash black line).
No continuum normalization was made. Smooth blue gaussian mimic position and intensity of expected C$_{60}^+$ feature (normalized to the 9577~\AA\ intensity).
}
\label{9428area}
\end{figure}

Let's compare the discovered DIB profiles after the continuum normalization. This is demonstrated
in Fig. 3 where the normalized spectra are overplotted on the laboratory spectrum model calculated
according to the most recent version of C$_{60}^+$ data (Campbell et al. 2016). For each target the model spectrum is normalized to the observed
intensity of 9577~\AA\ band. Thus the Fig. 3 demonstrates the relative intensities of four bands, attributed
to C$_{60}^+$. The 9428~\AA\ feature remains below
the level of detection. We also measured the rest wavelengths of the considered DIBs (Table 1).
Let's emphasize also that the whole spectral range, depicted in Figs. 2 and 3, is full of
weak stellar lines and it is thus difficult to separate reliably possible stellar contaminations from
interstellar bands. The best argument is that all interstellar features should be of the same
(more or less) profiles, doesn't matter what's the rotational velocity of the observed star.
Stellar lines should be broadened and extended in spectra of the heavily reddened, rapidly rotating targets.

The fact that the strength ratio 9577/9366 is variable (Fig. 3, Table 1)  is a strong argument
against their common origin. Moreover, we are not able to detect the 9428~\AA\ band which should be
stronger than 9366~\AA, according to the laboratory data. Moreover, we have to face the problem,
presented already by Walker et al. (2015), i. e. that of the shape of the 9366 feature. As
mentioned above, there is a possibility that we observe two DIBs, partially blended, which
do not share the same carrier. Another possibility is that the carrier is centrosymmetric
and one of the observed peaks is just the bandhead. This could, for example, follow higher
or lower rotational temperatures of the DIB carrier, possibly correlated with that of
interstellar $C_2$. However, such a comparison is not possible. In BD +40 4220 the $C_2$
lines are single, in CygOB2\,12 they show evident Doppler splitting; which of the peaks is related
to DIBs? In HD183143 the $C_2$ is hardly detectable. It creates the question whether very
complex carbon molecule can be present while the simplest one is nearly absent? Also the abundance of
$C_2$ is at least twice as big in Cyg OB2\,12 as in BD +40\,4220. The ESPADONS spectra do not
allow to search for the violet $C_3$ because of high noise level in this range. The very weak
feature, seen in the laboratory spectra near 9348.4~\AA\, remains below the level of detection in all our spectra.

The results of our measurements are given in Table 1. The 9366~\AA\ feature was divided into two
possible bands. It is interesting that even if to take into account only its stronger
(in HD183143) component, it appears too strong (in relation to 9577~\AA) in most of the targets if the laboratory intensities
are applied. If the feature is not a blend, the EW(9366) intensity is
always too high. Let's emphasize that the 9366~\AA\ feature is very shallow and thus its measurements
-- difficult, because of high noise contamination. It seems of basic importance to check the
9366/9577 relation using a statistically meaningful sample but the latter must contain heavily
reddened objects as in other ones the considered features (especially 9366) are very weak and thus -- their measurements
uncertain.

Our very poor statistics (Table 1) allows to infer a vague conclusion that the intensities of infrared DIBs
are in a way correlated with those of other DIBs. This, however, does not bring any substantial information.
The sightlines to our heavily reddened targets intersect several clouds -- see Fig.  1. In such a case we
observe always a kind of ill--defined average and it is not astonishing that all interstellar features
are in such the averages correlated.

\section{Conclusions}

We have concluded in the previous paper -- Galazutdinov et al. (2017) -- that the attribution of
the two evident DIBs: 9632~\AA\ and 9577~\AA\ to $C_{60}^+$ is premature. The proper elimination of both telluric and stellar
contaminations -- necessary to check the rest wavelengths and strength ratios -- is very difficult. In the former
paper we have demonstrated that the 9632/9577 strength ratio is variable; now we show that 9577/9366 ratio is
variable as well. Moreover, the expected at 9428~\AA\ feature seems to remain below the level of reliable detection,
despite the fact that we tried to find it in the spectrum of Cyg OB2 12 -- the one of most heavily reddened stars ever
observed in high resolution. This is another argument against $C_{60}^+$ as the carrier of the considered bands
because the 9428~\AA\ band should be stronger than 9366~\AA\, according to the laboratory predictions.
We were not able to detect something resembling 9366 DIB in our sample of 19 objects (Galazutdinov et al. 2017) except, very vaguely,
three objects. The spectra, used in this paper allow a more precise elimination of telluric lines because the instrument
is situated on a much higher elevation which makes telluric features less saturated. It should also be emphasized that the strongest 9577 band is
in practically all objects broader than the laboratory band of $C_{60}^+$ (Fig. 3). The effect cannot be explained by the Doppler splitting as in
three objects the interstellar K{\sc i} line is narrow and, even if Doppler--splitted, cannot explain the observed broadening
of the feature. In general the statistics of high quality spectra of heavily reddened stars in the infrared range is poor and its extension
seems to be really important. However, the probability that $C_{60}^+$ is present in a substantial amount in
translucent interstellar clouds, being responsible for the observed features, is very low.

\Acknow{
GAG acknowledges the support of Russian Science Foundation (project 14-50-00043, area of focus Exoplanets) for support of
experimental part of this work.
JK acknowledges the financial support of the Polish National Science Center during the period 2015--2017 (grant 2015/17/B/ST9/03397).
This research used the facilities of the Canadian Astronomy Data Centre (ESPADONS data archive) operated by the
National Research Council of Canada with the support of the Canadian Space Agency.
}

\end{document}